\documentclass{amsart}

\usepackage{amssymb,amsfonts,latexsym,amsmath,amsthm,wrapfig,graphicx, hyperref, %showlabels
}

\usepackage{bm}
\usepackage{epsfig}

\input xy
\xyoption{all}

\numberwithin{equation}{section}

\newtheorem{remark}{Remark}

\newcommand{\be}{\begin{equation}}
\newcommand{\ee}{\end{equation}}
\newcommand{\la}{\label}

\begin{document}

\title[Cantilever tip dynamics]{A mathematical model for droplet separation by surface tension using contact cantilevers - applications to {\it{in situ}} diagnosis and treatment}

\date{May 2024. {\it{Contact:}} soniaelizabeth.teodorescu@gmail.com}

\author[Sonia Elizabeth Teodorescu]{Sonia Elizabeth Teodorescu  \\ Advisor: Leslaw Skrzypek}
%\email{soniaelizabeth.teodorescu@gmail.com}
%\address{4202 E. Fowler Ave., CMC342, Tampa, FL 33620}

\begin{abstract}
This work provides an exact mathematical characterization of the meniscus formed by a liquid of density $\rho$ (model for tumor tissue) when probed with a cantilever device, operating by gravity (acceleration $g$) and with surface tension coefficient $\sigma$ (material-dependent for the specific choice of liquid and cantilever). The shape and extremal parameters (maximum height $\mathcal{H}$, break-off volume $\mathcal{V}$) of the meniscus formed, as functions of $\sigma, \rho$, are found by an exact analysis. Having knowledge of the explicit relationship between these parameters allows to perform in one procedure both diagnosis  and treatment.% (using $\mathcal{D}$).
\end{abstract}

\maketitle

\section{Introduction}

A  slab of vertical plane section width $2 W$ is placed on top of a liquid of surface tension $\sigma$ and density $\rho$ and raised vertically so that the meniscus formed has height $H$ relative to the free surface of the liquid. The maximum contact angle between the liquid and the slab is $\vartheta$ ($\vartheta \to 0^+$ for the limit of perfect wetting). We wish to find the height $H_0$ at which the meniscus has minimum width, the minimum width of the meniscus, $W_0$, and the volume of liquid contained between the slab and the minimum-width horizontal section, as functions of the parameters $\sigma, \rho$. Furthermore, we use the general expressions to derive the maximum height of the meniscus, $\mathcal{H}$, for which $W_0 = 0$ and the surface breaks off, forming a droplet attached to the slab. Corresponding to this limit, we compute the break-off droplet volume $\mathcal{V}$ of the droplet and  maximum contact angle, $\vartheta$. The explicit formulas relating the measured quantities $\mathcal{H, V}$ to the computed parameters $\vartheta, \sigma/\rho$ form the basis for an efficient algorithm that can be implemented by this minimally-invasive approach.  These solutions allow to provide a quantitative procedure through which the cantilever use in medicine can be incorporated to provide both {\it{in situ}} investigative options (as the dependence $\frac{\sigma}{\rho}(\mathcal{H, V} )$ allows to immediately classify the probe as normal tissue or not) and treatment options, by targeted delivery of medication. 

\section{Acknowledgments}

The author wishes to thank the Harvard School of Medicine and the Summer 2023 Moonshot internship program for high-school students, for the opportunity to participate and for the invaluable immersive learning experience the program offered. The main question addressed in this article was first formulated during the 2023 Moonshot program, as the author was working towards the Capstone project and its final presentation. 

\section{The mathematical model}

\subsection{General analysis of the surface tension model}

 What follows refers to points on the free surface of the liquid (the meniscus) and is illustrated in Figure~1. Let $z$ denote the vertical position measured downwards (along the orientation of gravity and towards the free surface of the liquid), starting from the location of the slab. Then the hydrostatic equilibrium condition, which consists of the balance between the hydrostatic pressure and the pressure difference due to curvature of the free surface of the liquid (Laplace's Law), reads

\begin{equation}{\label{Eq}}
%$$
\rho g z = \sigma (\kappa_v + \kappa_h),
%$$
\end{equation}

\noindent where 
$$
\kappa_v = \frac{d \varphi}{d \ell} = \frac{\ddot x \dot z - \ddot z \dot x}{[(\dot z) + (\dot x)^2]^{\frac{3}{2}}} =  
\frac{\ddot x} {[1 + (\dot x)^2]^{\frac{3}{2}}}
$$

\bigskip

\noindent is the curvature of the meniscus in the vertical plane, with $x(z)$ denoting the horizontal coordinate measured downwards  from the center of the slab, 

\begin{equation}{\label{BV}}
%$$
x: [0, H] \to [W_0, \infty), \quad x(0) = W, \quad \lim_{z \to H^{-}} x(z) = \infty,
%$$
\end{equation}

\noindent and $\dot x$ denotes the derivative with respect to $z$, satisfying the extrema conditions 

\begin{equation}{\label{Extrema}}
\dot x(H - H_0) = 0, \,\, x(H - H_0) = W_0
\end{equation}
 
\noindent at the height $H_0$ where the meniscus reaches its minimum width, $W_0$. The angle $\varphi$ is the angle made by the tangent line to the curve $-z(x)$ with respect to the horizontal direction and $d \ell$ is the arclength differential. It follows from $\tan \varphi = - \frac{dz}{dx}$ that $\dot{x} = -\cot \varphi$. The second principal curvature, 

$$
\kappa_h = - \frac{1}{R_{\parallel}} < 0
$$

\noindent corresponds to the horizontal section through the meniscus and is completely determined by the shape of the slab, having opposite sign (convexity) relative to the vertical section curvature. There are two typical cantilever probe  geometries: cylindrical (for which the horizontal section through the meniscus, at any height, is a circle of radius $R_{\parallel} \le W = x(0)$) and parallelipipedic (with rectangular section, $R_{\parallel} \to \infty$), which will be the case treated in this analysis.

\begin{figure}[h!] 
\begin{center}
\includegraphics*[width=12cm]{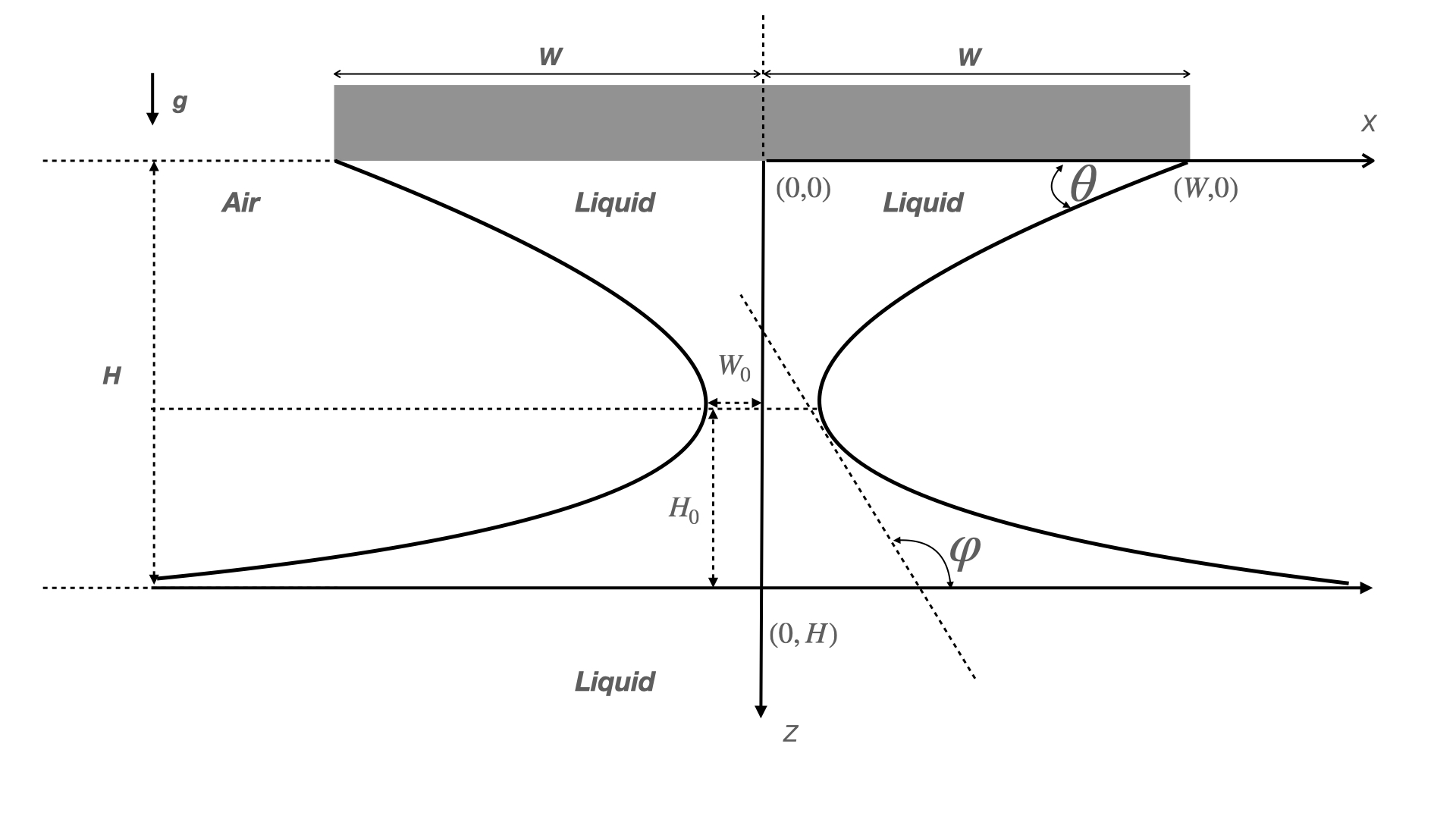}
\caption{The exprimental set-up diagram (shown in the vertical section). The coordinates used are $(x, z)$ (with the positive $z$ orientation taken along the gravitational acceleration direction).}
\label{fig1}
\end{center}
\end{figure}

\subsection{The case of a very long rectangular slab (in the horizontal section)}

In the following, we consider the  rectangular slab geometry $2W \times L$, for which there is only one non-vanishing principal curvature, as  $R_{\parallel} \sim L \to \infty$. (Then the Gaussian curvature of the meniscus $\kappa = \kappa_v \cdot \kappa_h \to 0$ and the surface is almost flat.) 

Let $\theta$ be the angle made by the tangent to the meniscus with the horizontal direction, at the boundary between slab, liquid, and air.  This quantity is the {\it{wetting angle}}, and it can take values between $0$ (for perfect wetting) and a maximum value, $\theta = \vartheta$, which is specific to the pair liquid-slab used. Since the function $\dot x$ is monotonically increasing, vanishing at $z_* = H - H_0,  \,\, \dot x(z_* = H - H_0) = 0$, and has endpoint limits $\dot x(z \to 0^+) \to -\cot(\theta), \,\,  \dot x(z \to H^-) \to -\cot(\pi^{-}) = \infty$, we can integrate the function $z^2(\dot{x})$ in \eqref{Eq} and we find 

$$
z^2(\dot x) = \frac{2 \sigma}{\rho g}  \int_{-\cot(\theta)}^{\dot x} \frac {d u} {[1 + u^2]^{\frac{3}{2}}} = 
\frac{2 \sigma}{\rho g} \sin(\tan^{-1}(u)) \Big{|}_{-\cot(\theta)}^{\dot x}, 
$$

\bigskip 

\noindent yielding the exact solution 

\begin{equation}{\label{Sol}}
%$$
z^2(\dot{x}) = \frac{2 \sigma}{\rho g} \left [\cos \theta + \frac{\dot x}{\sqrt{1 + \dot x^2}} \right ]
%$$
\end{equation}

%\bigskip

\begin{remark} \label{rem1}
It is important to note that the second-order ordinary differential equation \eqref{Eq} with initial-value conditions \eqref{BV} 
fully determine the solution for the entire interval $z \in [0, H]$. Therefore, the limiting value of the derivative, $\dot{x}(z \to 0^+) = -\cot(\theta)$, and by extension, the contact angle $\theta$, are fully determined by fixing the independent variable $H \in [0, \mathcal{H}]$. In that sense, the entire derivation presented here is functionally dependent on $H$ and parametrically dependent on $W$. 
\end{remark}

\noindent 
At the height $z_* = H - H_0$, where the meniscus width reaches its minimum, $\dot{x} = 0$ and the solution \eqref{Sol} becomes
$$
(H-H_0)^2 =   \frac{2 \sigma}{\rho g}  \cos \theta 
%\int_{-\infty}^{0} \frac {d t} {[1 + t^2]^{\frac{3}{2}}} =  \frac{2 \sigma}{\rho g} 
%\Rightarrow \rho g (H-H_0) = \frac{\sigma}{\frac{H-H_0}{2\cos \theta}}, %= \sigma \kappa_0, 
$$
%\noindent which shows (by \eqref{Eq}) that the curvature of the meniscus at its minimum width equals 
%$
%\kappa_* = \frac{2\cos \theta}{H-H_0}.
%$
%Taking the maximal domain of integration $\dot{x} \in [-\cot \theta, \infty)$, we obtain 
Equation \eqref{Sol} also allows to express the total height of the liquid column by taking $\dot{x} \to \infty$, which corresponds to the free surface of the liquid: 

\begin{equation}{\label{Max}}
%$$
H^2 =  \frac{2 \sigma}{\rho g} \left [1 + \cos \theta \right ]
%\int_{-\infty}^{\infty} \frac {d t} {[1 + t^2]^{\frac{3}{2}}} =  \frac{4 \sigma}{\rho g} = 2(H-H_0)^2
\Rightarrow \left (\frac{H-H_0}{H}\right )^2 = \frac{\cos \theta}{1 + \cos \theta}.
%$$
\end{equation}
%
%Therefore,
%$$
%\kappa_* = \frac{2 \cos \theta}{H-H_0} = \frac{2\sqrt{(1+\cos \theta) \cos \theta}}{H}. %\quad 
%$$
%
%\bigskip

As mentioned in Remark~\eqref{rem1}, we notice that all the quantities of interest are proportional to an arbitrary common scale factor, which can be taken to be $H$ or, as we will see in the next section, the half-width of the slab, $W$. %In particular, the depth at which the meniscus reaches its minimal width, $z_*$, depends only on the material parameters $\sigma, \rho, \theta$, so that the break-off depth is actually independent of $W$, and equals
Explicitely, the geometric parameters of the solution, $H, H_0, \theta$ are related by the formulas

\begin{equation}{\label{Parameters}}
H^2 =  \frac{2 \sigma}{\rho g} \left [1 + \cos \theta \right ], \quad \left (\frac{H-H_0}{H}\right )^2 = \frac{\cos \theta}{1 + \cos \theta}, 
\end{equation}

\noindent which means that for the extremal shape (when $H, H_0$ reach their maximal values),

\begin{equation}{\label{Extremal}}
%$$
\mathcal{H - H}_0 = \sqrt{\frac{2 \sigma}{\rho g}  \cos \vartheta}, \quad \mathcal{H} =  \sqrt{\frac{2 \sigma}{\rho g} \left [1 + \cos \vartheta \right ]}.
%$$ 
\end{equation}

\subsubsection{An algorithm for finding material parameters by non-invasive measurements}

It may appear tempting to use formulas \eqref{Extremal} to infer the values of $\vartheta, \sigma/\rho$, but that is deceiving. For a minimally-invasive cantilever device, measuring $\mathcal{H}$ is possible, but $\mathcal{D} \equiv \mathcal{H} - \mathcal{H}_0$ is not directly accessible. Therefore, we need to find another expression relating $\mathcal{H}$ to $\vartheta$, parametrically-dependent on the size of the slab, $W$, so that we can devise an effective algorithm for retrieving the quantity $\sigma/\rho$ directly from measuring only $\mathcal{H}$ and the sectional area at pinch-off, $\mathcal{A}$.

In order to find the maximal possible height  $\mathcal{H}$ attained by the meniscus, we will solve \eqref{Eq} for $\dot{x}(z)$ and require that the minimum width $x(\mathcal{D})$ vanish, or

\begin{equation} \label{height}
%$$
x(\mathcal{D}) - x(0)  = 0 - W = \int_0^{\mathcal{D}} \dot{x} dz
%$$
\end{equation}

%Denoting the material constant 
%
%$$
%\frac{2 \sigma \cos \theta}{\rho g}  = (H-H_0)^2 = H^2 \frac{\cos \theta}{1 + \cos \theta} \equiv \mathcal{D}^2
%$$

\noindent This is equivalent to requiring $W_0 = 0$, or the limit of the liquid developing a singularity at its minimal width section. Further raising the slab would then produce a pinch-off separation of the liquid and the meniscus will break, having reached the maximal possible height.  Introducing the new variable 

$$
\lambda \equiv \frac{z^2 - \mathcal{D}^2}{\mathcal{D}^2} \in \left [-1, \frac{1}{\cos \vartheta} \right ], \quad 
\lambda(0) = -1, \quad \lambda(\mathcal{D}) = 0,
$$

%\bigskip

\noindent we arrive at the equation 

$$
\dot{x} = \frac{\lambda \cos \vartheta}{\sqrt{1-\lambda^2\cos^2\vartheta }} 
%\Rightarrow \dot{x} = \frac{d x}{d \lambda} \cdot \frac{d\lambda}{dz}=  \frac{2z}{\mathcal{D}^2} \cdot  \frac{d x}{d \lambda} 
\Rightarrow 
\frac{d x}{d \lambda} = \frac{\mathcal{D}}{2} \frac{\lambda \cos \vartheta }{\sqrt{(1+ \lambda)(1-\lambda^2 \cos^2\vartheta)}}
$$

%\bigskip

Using \eqref{Extremal}, the maximum height condition \eqref{height} becomes

%$$
%\int_{-1}^0 \frac{\mathcal{D}}{2} \frac{\lambda}{(1+ \lambda)\sqrt{1-\lambda}} d\lambda = -W \Rightarrow 
%$$
%\bigskip

\begin{equation}\label{int}
\frac{W}{H} \sqrt{\frac{1+\cos \vartheta}{\cos \vartheta}} = 
\int_{0}^1 \frac{\lambda \cos \vartheta }{\sqrt{(1- \lambda)(1-\lambda^2 \cos^2\vartheta)}}
 d\lambda
\end{equation}

\bigskip

%\noindent 
The right-hand side is expressed as the generalized hypergeometric function \cite{Slater}

$$
\int_{0}^1 \frac{\lambda \cos \vartheta }{\sqrt{(1- \lambda)(1-\lambda^2 \cos^2\vartheta)}} d\lambda = \frac{4\cos \vartheta}{3}  {}_3F_2 \left (\frac{1}{2}, 1, \frac{3}{2};  \frac{5}{4}, \frac{7}{4}; \cos^2 \vartheta \right ),
$$

\bigskip
\noindent so we can conclude that, for a slab of total width $2W$, the maximum height at which the slab can be raised, above the free surface of the liquid, at which the meniscus breaks and a droplet is formed, equals 

\bigskip

\begin{equation}\label{formula}
%$$
\mathcal{H} = \frac{3}{4} \sqrt{\frac{1+\cos \vartheta}{\cos^3 \vartheta}}\cdot  \frac{1}{ {}_3F_2 \left (\frac{1}{2}, 1, \frac{3}{2};  \frac{5}{4}, \frac{7}{4}; \cos^2 \vartheta \right )} \cdot W \equiv \mathcal{F}(\vartheta) W.
%$$
\end{equation}

\bigskip
\noindent In \eqref{formula} we have introduced the function $\mathcal{F}(\vartheta)$, independent of any parameters of the problem, so the extremal wetting angle $\vartheta$ is found from 
$
\vartheta = \mathcal{F}^{-1} \left ( \frac{\mathcal{H}}{W}\right).
$

\begin{remark}
The integral in \eqref{int} diverges logarithmically  in the limit $\vartheta \to 0^+$, i.e. 
$$
\lim_{\epsilon \to 0^+}
\int_{0}^{1-\epsilon} \frac{\lambda }{1- \lambda}
 d\lambda  =
 \lim_{\epsilon \to 0^+} [\epsilon - 1 - \ln \epsilon ] = \infty 
 \Rightarrow \lim_{\vartheta \to 0^+} \mathcal{F}(\vartheta) = 0.
$$
\noindent This analysis indicates that perfect wetting at break-off, $\vartheta \to  0^+$, can only be achieved in either the case of an infinitely-wide slab, when $W \to \infty, \mathcal{H} \ne 0$, or through break-off right on contact (with no meniscus formation), that is for $\mathcal{H} = 0$.
\end{remark}

\subsection{An algorithm for minimally-invasive investigation via cantilever tips}

Formulas \eqref{Extremal}, \eqref{formula} provide us with an explicit method for extracting the material parameters $\vartheta, \sigma/\rho$ directly from the measurement of the maximum height, $\mathcal{H}$. We recall that the half-width of the slab, $W$, is known as a controlled variable. From \eqref{formula}, we find the unique solution $\cos \vartheta \in [0, 1)$ (as $\vartheta \ne 0$ for $\mathcal{H} \ne 0, W$ finite) as

\begin{equation} \label{hyper}
%$$
  {}_3F_2 \left (\frac{1}{2}, 1, \frac{3}{2};  \frac{5}{4}, \frac{7}{4}; \cos^2 \vartheta \right ) \sqrt{\frac{\cos^3 \vartheta}{1+\cos \vartheta}}=  \frac{3 W}{4 \mathcal{H}} 
%$$
\end{equation}

%\bigskip

\noindent With this value for $\cos \vartheta$, expression \eqref{Extremal} then provides the explicit formula

\begin{equation} \label{final}
\frac{\sigma}{\rho} = \frac{g \mathcal{H}^2}{2(1 + \cos \vartheta)}. 
\end{equation}

Let us now compute the area of the vertical section of the droplet at pinch-off, by using the symmetry shown in Figure~1 and  writing 

$$
\mathcal{A} = 2 \int_0^W z dx =  -2 \int_0^{\mathcal{D}} z \dot{x} dz = - \int_0^{\mathcal{D}^2} \dot{x} dz^2
$$

\noindent where we have made use of the fact that the functions $x(z), \dot{x}(z)$ are strictly-monotonically decreasing on the interval $z \in [0, \mathcal{D}]$. 

Formula \eqref{Sol} allows to solve for $\dot{x}$ as a function of $z^2$ as

$$
z^2 - \mathcal{D}^2 = \frac{2\sigma}{\rho g} \frac{\dot x}{\sqrt{1 + \dot{x}^2}} \Rightarrow
\dot{x} = \frac{z^2 - \mathcal{D}^2}{\sqrt{\left(\frac{2\sigma}{\rho g}\right)^2 - \left (z^2 - \mathcal{D}^2 \right )^2}}
$$

For simplicity, we introduce the new substitution variable 

$$
\nu \equiv \frac{\rho g}{2\sigma} ( \mathcal{D}^2 - z^2), \quad \nu \in [0, \cos \vartheta], \quad \frac{dz^2}{d\nu \,\,} = -\frac{2 \sigma}{\rho g},
$$

\noindent whose range was found from \eqref{Extremal},  and write the area as 

$$
\mathcal{A} = \frac{2\sigma}{\rho g} \int_0^{\cos \vartheta} \frac{\nu}{\sqrt{1-\nu^2}} d\nu = -  \frac{2\sigma}{\rho g} \sqrt{1 - \nu^2} \Big{|}_{0}^{\cos \vartheta}, 
$$

\noindent yielding the final result

\begin{equation} \la{Area}
\mathcal{A} = \frac{2 \sigma}{\rho g} (1-\sin \vartheta) 
\end{equation} 
\bigskip

Together with the extremal height at pinch-off obtained in \eqref{Extremal}, this leads to the system of equations

\begin{equation} \la{System}
\left \{
\begin{array}{lll}
\mathcal{A} & = & \frac{2 \sigma}{\rho g} (1-\sin \vartheta) \\
&& \\
 \mathcal{H}^2 &= &  \frac{2 \sigma}{\rho g} \left [1 + \cos \vartheta \right ]
\end{array}
\right .
\end{equation}

\noindent which can be solved by elementary methods to obtain the explicit, exact solution

\begin{equation} \la{Angle}
\vartheta = 2 \arctan \left [1 - \sqrt{\frac{2 \mathcal{A}}{\mathcal{H}^2}} \right ]
\end{equation}

Using trigonometric identities in \eqref{System}, this also leads to the final exact result 

\begin{equation} \la{ExactFinal}
\frac{\sigma}{\rho g} = \frac{\mathcal{H}^2}{4}
\left [ 1 + 
 \left (1 - \sqrt{\frac{2 \mathcal{A}}{\mathcal{H}^2}} \right )^2
 \right ]
\end{equation}

\begin{remark}
As previously indicated, the only measurements required are those of the maximal height of the meniscus, 
$\mathcal{H}$ and pinch-off droplet volume $\mathcal{V}$,  obtained by a non-invasive implementation of the  probe. The maximal height measurement is obtained by the direct observation of the vertical displacement of the cantilever tip, between making contact with the free surface of the liquid and the height at pinch-off.  The area of the vertical section of the droplet, $\mathcal{A}$,  is found from  the volume $\mathcal{V}$ of the pinched-off droplet, which by incompressibility must equal $\mathcal{V} = L \cdot \mathcal{A}$, for a rectangular horizontal section of the slab  with (known) dimensions $2W \times L$.  %Therefore, this procedure allows for an exact computation of the material-specific ratio $\sigma/\rho$, by means of a simple, direct measurement.  
\end{remark}

%\newpage 

\subsubsection{Numerical evaluations and applications to effective discriminating analysis}

Notice that equations \eqref{System} imply the inequality 

$$
\frac{2 \mathcal{A}}{\mathcal{H}^2} = 2 \frac{1 - \sin \vartheta}{1 + \cos \vartheta} \le 1, \,\, \vartheta \in \left [0, \frac{\pi}{2} \right ],
$$

\bigskip

\noindent as this function of $\vartheta$ has negative derivative and takes values between 1 at $\vartheta = 0$ and 0 at $\vartheta = \frac{\pi}{2}$. The exact value of the ratio $\sigma/\rho g$ has therefore the sharp theoretical bounds  $\frac{\mathcal{H}^2}{4}, \frac{\mathcal{H}^2}{2}$, 
%$$
%\frac{\mathcal{H}^2}{4} \le \frac{\sigma}{\rho g} \le \frac{\mathcal{H}^2}{2},
%$$
%\noindent 
or in asymptotic notation \cite{Landau} $\sigma/\rho g = \Theta(\mathcal{H}^2)$. Taking a numerical example, we compare the values of $\sigma/ \rho g$ between blood and mercury: at standard temperature and pressure (STP), we obtain  \cite{Data1},  \cite{Data2}
$$
\left ( \frac{\sigma}{\rho g} \right )_{H_2O |_{STP}} = 5.692 \times 10^{-6} m^2, \quad \left ( \frac{\sigma}{\rho g}  \right )_{Hg|_{STP}} = 3.594 \times 10^{-3} m^2,
$$

%\bigskip

\noindent and as the two values differ by 3 orders of magnitude, this indicates that the two liquids can be differentiated live  though this probing procedure and furthermore, by obtaining an estimate for the order of magnitude of $\mathcal{H}$, abnormalities can be detected and identified efficiently.  As we elaborate in \S \ref{Conclusions}, this has important practical applications, potentially leading to significant reduction in treatment timeframes. 

\newpage

\subsubsection{Approximations by elementary functions in the limiting case $1 - \cos \vartheta \to 0$}

For a generic value of the maximum wetting angle $\vartheta > 0$, we can use the power series expansion \cite{Slater} for the hypergeometric function which allows approximating the solutions with arbitrary precision:

$$
 {}_3F_2 \left (\frac{1}{2}, 1, \frac{3}{2};  \frac{5}{4}, \frac{7}{4}; \cos^2 \vartheta \right ) = 
 1 + \frac{12}{35}\cdot {\cos^2 \vartheta}+ \frac{8}{77} \cdot {\cos^4 \vartheta}+\cdots
$$

\noindent Of particular interest is the case of near-perfect wetting at break-off, when $\vartheta$ is very small, $\vartheta \to 0^+$
%so we can approximate
%
%$$
%\cos \vartheta = 1 - \frac{\vartheta^2}{2} + \ldots, \quad \cos^2 \vartheta = 1 - \vartheta^2 + \ldots, %\quad  {}_3F_2 \left (\frac{1}{2}, 1, \frac{3}{2};  \frac{5}{4}, \frac{7}{4}; \cos^2 \vartheta \right ) = \ldots 
%$$
%
%\bigskip 
%
and we can use the expansion  \cite{Slater} valid for $|\vartheta| \ll 1$, to order $o(|\vartheta|)$ \cite{Landau}

$$
{}_3F_2 \left (\frac{1}{2}, 1, \frac{3}{2};  \frac{5}{4}, \frac{7}{4}; 1 - \vartheta^2 \right ) \to 
-4 \frac{\Gamma \left ( \frac{5}{4}\right )\Gamma \left ( \frac{7}{4}\right )}{\Gamma^2\left ( \frac{1}{2}\right )} \ln \vartheta + c = - 3\sqrt{2} \ln \vartheta + c,
$$

%\bigskip

\noindent where $\Gamma$ is Gauss' Gamma function and $c$ is a constant, given explicitly by 

$$
c = - \frac{\sqrt{3}}{2} \left [\psi\left (\frac{1}{2} \right ) + \psi \left ( 1 \right ) + 2\gamma + \frac{1}{8} {}_4F_3  \left (\frac{7}{8}, \frac{9}{8}, 1, 1; \frac{3}{2}, 2, 2; 1 \right )\right ],
$$

\bigskip
\noindent with $\psi(z), \gamma$ the digamma function and the Euler-Mascheroni constant, respectively.  Plugging into \eqref{hyper}, we obtain the non-perturbative expansion 

$$
\vartheta \simeq A e^{-\frac{W}{4\mathcal{H}}}, \quad \vartheta \to 0^+,
$$

\noindent where we introduce the notation

\begin{equation} \label{A}
%$$
A = e^{\frac{\sqrt{2} \, c}{2}} %\simeq 1.307562... 
%$$
\end{equation}
so  in \eqref{final}, this approximation leads to

$$
\frac{\sigma}{\rho} = \frac{g \mathcal{H}^2}{4} \left [1   + \frac{A^2}{4} e^{-\frac{W}{2\mathcal{H}}} \left (1 + \ldots \right ) \right].
$$

\bigskip
\noindent Therefore, the expansion in the $ |\vartheta| \ll 1, \mathcal{H} \ll W$ limit is:

\begin{equation} \label{algorithm}
%$$
%\left (\frac{\sigma}{\rho} \right )_0 = \frac{g \mathcal{H}^2}{4},  \quad 
\frac{\sigma}{\rho}  \simeq \frac{g \mathcal{H}^2}{4} \left [1   + \frac{A^2}{4} e^{-\frac{W}{2\mathcal{H}}} \right]
%\frac{g \mathcal{H}^2}{4} \left [1 + \frac{1}{4} e^{-\frac{3\sqrt{2}}{4A}\frac{W}{\mathcal{H}}}\right] 
= \frac{g \mathcal{H}^2}{4} \left [1 + \frac{e^{\sqrt{2} c}}{4} e^{-\frac{W}{2\mathcal{H}}}\right],
%$$
\end{equation}

\noindent which demonstrates that the limit of near-perfect wetting at pinch-off (corresponding to $\vartheta \to 0^+$) is non-perturbative and exponentially accurate, in the sense that the $0^{\rm{th}}$ order result, 

$$
\frac{\sigma}{\rho} \Big{|}_{\vartheta \to 0^+} = \frac{g \mathcal{H}^2}{4}, 
$$

\noindent holds up to pure exponential corrections (beyond all orders in approximation sense).

%\newpage 

\section{Applications and further possible extensions} \la{Conclusions}

The 2023 Moonshot program was designed around the challenge to design a multi-functional robotic investigative device which would allow for non-invasive inspection of the respiratory system, visual tumor detection, {\it{in situ}} diagnosis, and (potentially) local administration of medication or  other liquid-based treatments. One of the main obstacles for this approach is finding a way to quickly detect abnormal tissue in the course of the endoscopic exploration, without the need for a regular biopsy, in which a sample of cores extracted from the tumor would be sent to a lab for microscopic evaluation and the results would only be known after a significant amount of time. The other imaging technique, MRI-based, is also known to be associated with relatively long waiting time for patients. Overall, lung disease patients have an average referral-to-treatment time (identification to therapeutic resection) of 98 days, of which 47 days are associated with the investigations described here (referral to diagnosis) \cite{Report}. 

The mathematical model presented in this article provides the rigorous analysis for meniscus dynamics in a realistic cantilever geometry, appropriate for the real-life situation which is the focus of the Moonshot program. Using the explicit formulas  \eqref{Angle}, \eqref{ExactFinal} \eqref{algorithm}, it is possible to implement and deploy an effective approach for minimally-invasive {\it{in situ}} diagnosis, making use of an affordable robotic device which would greatly reduce the reliance on  specialized technology and procedures. The exponential sensitivity of the $\frac{\sigma}{\rho}$ ratio with respect to the measured quantity $\frac{W}{\mathcal{H}}$ lends this approach very high discriminating power, which allows it to provide very accurate identification of abnormal tissue, while remaining basically non-invasive (as only microscopic amounts would be involved in the course of droplet formation and break-off). By comparing the measured ratio $\sigma/\rho$  of  the sample to the known value for normal tissue, the procedure would provide an immediate classification of the sample being benign or not, which is another significant advantage. The ability to simultaneously diagnose and treat (e.g., by placing markers to be used in later stages of the diagnosis) would represent a significant advance over the approach currently in use for lung disease patients. 

In future work, we plan to provide a complete example for the deployment of these new techniques and a comprehensive analysis of the expected reduction in referral-to-treatment timeframes. 

%\vspace{5in}

\end{document}